\documentclass{article}

\usepackage{arxiv}

\usepackage[utf8]{inputenc} 
\usepackage[T1]{fontenc}    
\usepackage{hyperref}       
\usepackage{url}            
\usepackage{booktabs}       
\usepackage{amsfonts}       
\usepackage{cleveref}       
\usepackage{graphicx}
\usepackage[square,numbers,sort&compress,comma]{natbib}
\usepackage{indentfirst}
\usepackage{multirow}
\usepackage{siunitx}

\newcolumntype{P}[1]{>{\centering\arraybackslash}p{#1}}

\title{Investigating representation schemes for surrogate modeling of High Entropy Alloys}
\author{Arindam Debnath\\
Department of Materials Science and Engineering\\
Pennsylvania State University\\
University Park, PA 16802\\
\And
Wesley F.~Reinhart\\
Department of Materials Science and Engineering\\
Institute for Computational and Data Sciences\\
Pennsylvania State University\\
University Park, PA 16802\\
\texttt{reinhart@psu.edu}\\
}
\date{\today}

\begin{document}

\maketitle

\begin{abstract}
The design of new High Entropy Alloys that can achieve exceptional mechanical properties is presently of great interest to the materials science community. 
However, due to the difficulty of designing these alloys using traditional methods, machine learning has recently emerged as an essential tool.
Particularly, the screening of candidate alloy compositions using surrogate models has become a mainstay of materials design in recent years.
Many of these models use the atomic fractions of the alloying elements as inputs.
However, there are many possible representation schemes for encoding alloy compositions, including both unstructured and structured variants.
As the input features play a critical role in determining surrogate model performance, we have systematically compared these representation schemes on the basis of their performance in single-task deep learning models and in transfer learning scenarios.
The results from these tests indicate that compared to the unstructured and randomly ordered schemes, chemically meaningful arrangements of elements within spatial representation schemes generally lead to better models.
However, we also observed that tree-based models using only the atomic fractions as input were able to outperform these models in transfer learning.

\end{abstract}

\section{Introduction}
\label{sec:introduction}
High Entropy Alloys (HEAs) have been the subject of much interest since the concept was simultaneously and independently introduced by Cantor et al. \cite{Cantor2004} and Yeh et al.\cite{Yeh2004} in 2004.
Unlike conventional alloys, where small amounts of various elements are added to one principal element (e.g., bronze and steel), HEAs contain multiple principal elements, usually 5 or more, with molar fraction of each element between 5 and 35\% \cite{Cantor2004,Yeh2004,Senkov2018}. 
HEAs have been reported to show combinations of exceptional mechanical properties that are not attainable in conventional alloys, such as superior strength and room temperature ductility \cite{ Klimenko2021machine,Jung2021}. 
Particularly, refractory HEAs are of special interest as they are able to retain their mechanical properties at extreme temperatures, making them attractive for applications in future energy applications like gas turbines and jet engines.
Unfortunately, only a handful of the discovered HEAs have been able to surpass the performance of current generation of turbine materials.
Therefore, the design of new HEAs that can meet these operational requirements has become an area of intense scientific interest.

The design of HEAs with desired value of properties is a challenging task due to the large and mostly uncharted design space and the difficulty in predicting the non-linear relationships between structure, property, and processing parameters -- especially well outside the temperature range of conventional alloys.
To further complicate matters, the three traditional paradigms of materials design, namely the trial-and-error method, chemical intuition governed by physical and chemical rules, and computer simulations are also unsuitable for designing HEAs with target properties.
As such, Machine Learning (ML) tools such as high-throughput screening, Bayesian design of experiments, and deep-learning-based generative models have been used to accelerate the process of alloy design \cite{Klimenko2021machine,Wen2019, Yang2022,baird2022data, khatamsaz2022multi, giles2022machine,Bhandari2020}.
These models are able to leverage the learned patterns from the data provided to it to make predictions on unseen data.
ML surrogate models have especially become popular for materials design as they can be used to screen out potential candidates from a large pool as well as provide a baseline to compare the performance of deep learning models like Generative Adversarial Networks for inverse design \cite{Debnath2021}.
However, there are certain challenges that need to be addressed when it comes to utilizing ML tools for materials science problems, most of which arise because of adapting pre-existing methods to solve specific problems.

First, it is necessary to pre-process the raw data to a machine readable format while also preserving the relevant chemical information about each material.
Previous studies involving prediction of HEA phases or properties and generation of new HEA candidates have employed features that are derived from the alloy composition \cite{Jha2018Elemnet, Feng2021, Wen2019}.
Specifically, common representation schemes utilize the atomic fractions of the elements of the alloy, which may be in either an arbitrary or a structured manner following some chemical ordering or domain knowledge. 
For example, these structured representations can take the form of 1D vector where the elements are arranged according to their atomic numbers or a 2D matrix that follows the periodic table arrangement \cite{Jha2018Elemnet,Feng2021}.

Secondly, most traditional deep learning models use training datasets with $10^4 - 10^5$ observations, while most HEA datasets contain around $10^2 - 10^3$ datapoints.
As such, there is a serious risk of overfitting if deep learning models (which can easily have $10^4 - 10^6$ learned parameters) are trained directly on such small samples. 
Recently, Feng et al.\cite{Feng2021} proposed a ``General and Transferable Deep Learning'' (GTDL) framework for performing predictive tasks on small datasets using some structured 2D representation schemes.
They concluded that the 2D, periodic-table-inspired representation provided superior performance in transfer learning (TL), a ML technique wherein a representation learned by a model on one task is applied to achieve lower errors on another related task, often with fewer available data.
While they were able to successfully demonstrate the effectiveness of their chosen representation for one TL task, that study raised an important fundamental question: are 2D structured representations more effective than 1D structured or completely unstructured representations for HEA property prediction and design in general?

In this work, we investigated the impact of these different representation schemes when used as input on ML model performance.
Specifically, we have tried to address the following three research questions: 

Do any of these representation schemes (2D, 1D, unstructured) ...
\begin{enumerate}
    \item ... provide a definitive advantage when used for training a classification model?
    \item ... result in superior model performance when used for transfer learning on new domains? 
    \item ... provide better stability when performing transfer learning on especially small datasets?
\end{enumerate}

\section{Materials and Methods}

\subsection{Datasets}

Similar to the framework followed in Ref.~\citenum{Feng2021}, we use four different datasets for evaluating the different representation schemes on the three tasks specified in the previous section.
The frequency of the elements in these four datasets can be seen in Fig.~\ref{fig:element_counts}.
For details about these data beyond what is presented in the following subsections, readers are encouraged to refer to the original reports given in Refs.~\citenum{Feng2021, Yang2022, couzinie2018}.
\subsubsection{Glass Formation Ability (GFA)}

To first compare the different representation schemes on a particular deep learning task, GFA dataset, a medium-sized dataset originally containing 10,440 observations compiled by Feng et al. from multiple sources. 
While the original dataset spans 97 elements in the periodic table, the elements Rf, Db, Sg, Bh and Hs were removed as the Pettifor and modified Pettifor order do not account for these elements and we wished to train the different models on the same amount of data.
These elements only appeared in 5 observations (0.05\%).
Next, the dataset was expanded using the scheme discussed in Ref.~\citenum{Feng2021}, where the material phase (crystalline or amorphous) for each alloy in the dataset is resolved as a function of two different cooling rates $10^6 - 10^5 \mathrm{K/s}$ for melt spun, $10^2 - 10^0 \mathrm{K/s}$ for copper mold casting). 
This yielded a final training dataset of 20,870 observations.
    
\subsubsection{HEA Phase}

\begin{figure}
    \centering
    \includegraphics[width=\textwidth]{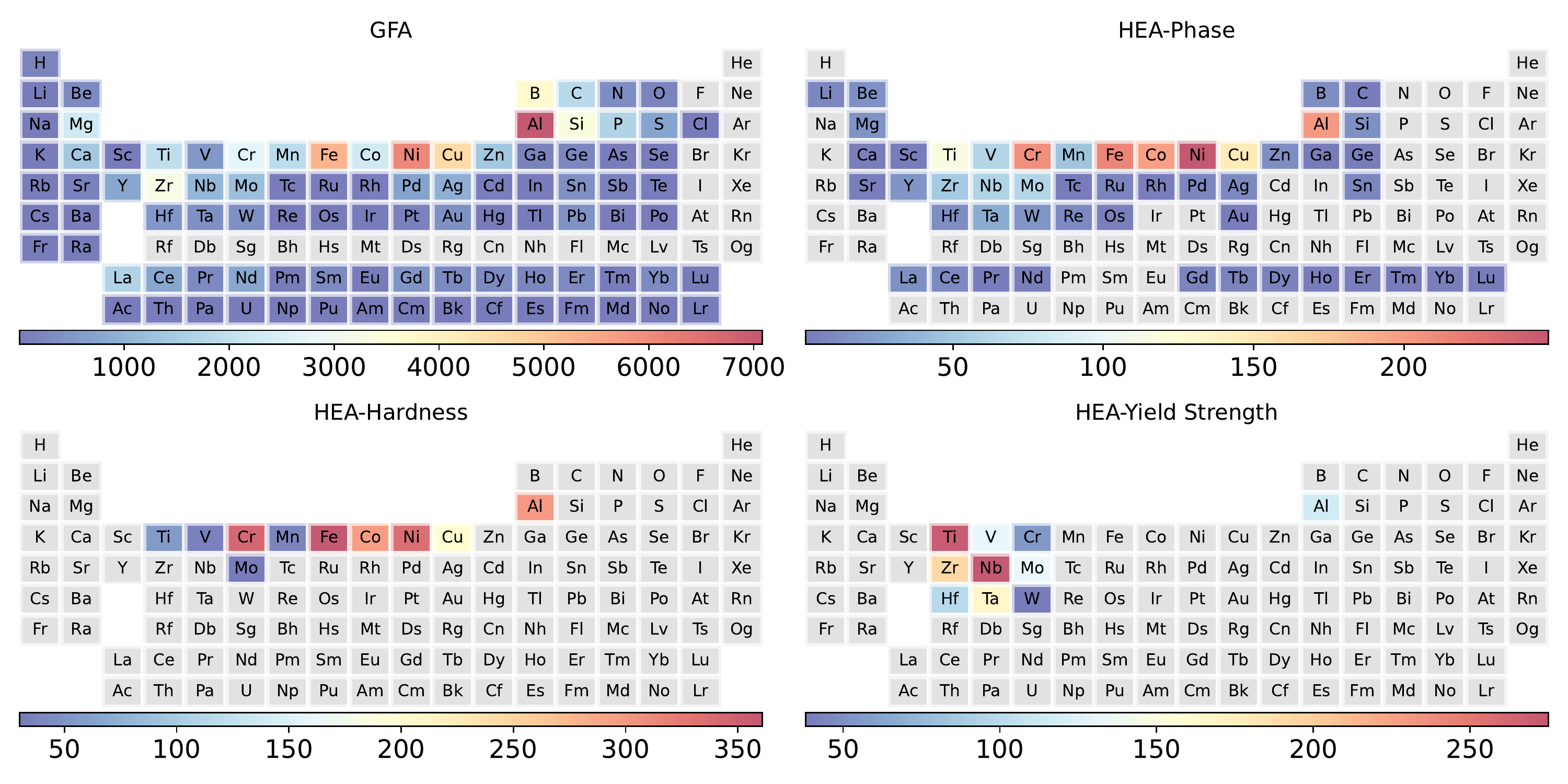}
    \caption{The frequency of the elements occurring in the GFA (top left), HEA Phases (top right), HEA Hardness (bottom left), and HEA Yield Strength (bottom right) datasets respectively.
    Compared to the GFA dataset, the HEA datasets have smaller data volume and more restricted domain in terms of elemental composition.}
    \label{fig:element_counts}
\end{figure}

The HEA Phase dataset is from Gao's review paper \cite{gao2017thermodynamics} and contains 355 experimentally synthesized HEAs and their reported phases. 
The number of elements in the HEAs in the dataset range from 2 to 9, with the median and mode both being 5 elements. 
There are 50 unique elements in the dataset, with frequencies shown in Fig.~\ref{fig:element_counts}. 
The majority of the alloys in the dataset are multi-phase (217), while single-phase solid solution HEAs like BCC, FCC and HCP account for 41, 24 and 14 observations respectively, with the remaining being amorphous.

\subsubsection{HEA Hardness}

Yang et al.\cite{Yang2022} compiled a HEA Hardness dataset from multiple sources, consisting of 370 HEA compositions made using vacuum arc melting and their corresponding as-cast Vickers hardness measured at room temperature.
The dataset contains four to seven components, with the median and mode both being 5 elements.
Notably, the dataset only covers 10 elements of the periodic table (Al, Co, Cr, Cu, Fe, Mn, Mo, Ni, Ti and V), a far departure from the GFA and the HEA phase datasets.

\subsection{HEA Yield Strength}

The HEA Yield Strength data used in this study were compiled by Couzini\'e et al. \cite{couzinie2018}.
While the original dataset contained 340 measurements for 120 unique compositions, we only include metallic elements -- removing any compositions containing C, S, O, or Si -- which resulted in a total of 300 measurements in the final dataset.
Apart from the alloy composition, the metallurical state at which the alloy was tested (as-cast or after optimization by annealing, hot isostatic pressured etc.), the phase content (single phase or combination of phases) and the testing temperature are also reported.
These additional features have been stated to influence the yield strength of the alloy \cite{Bhandari2020,giles2022machine}.

\subsection{Representation schemes and model architectures}

The different representation schemes discussed are closely tied to different model implementations, so they will be discussed in conjunction.
As described in Section~\ref{sec:introduction}, we consider different structuring schemes for the input data including completely unstructured, different 1D arrangements, and the 2D spatial structuring suggested by Ref.~\citenum{Feng2021}.

\begin{figure}
    \centering
    \includegraphics[width=\textwidth]{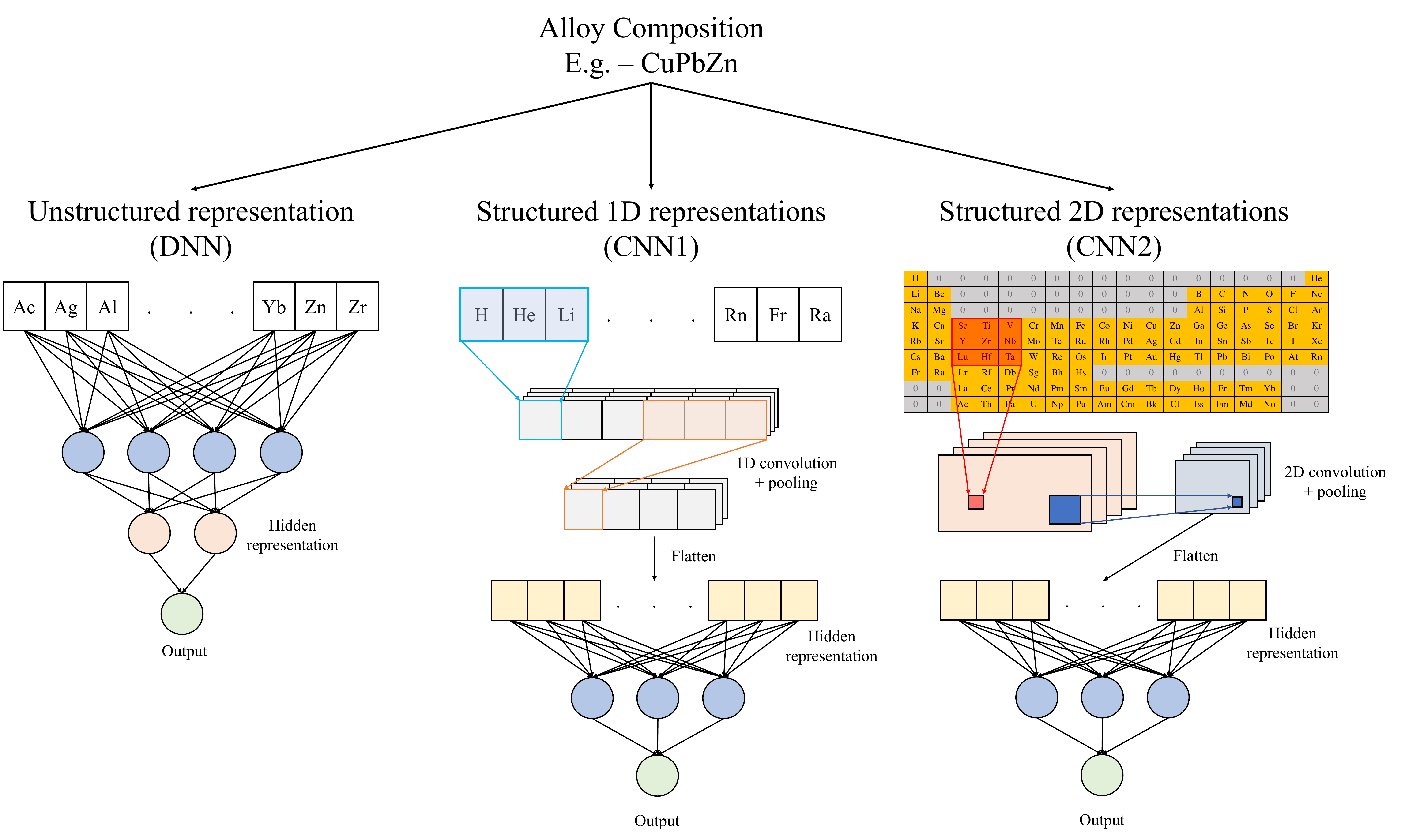}
    \caption{Schematic illustration of the different data representations and their associated model architectures.}
    \label{fig:representation_schemes}
\end{figure}

\subsubsection{Unstructured representation}

The unstructured representation consists of a vector array with dimension 103, with each component carrying the atomic fraction of one element from the periodic table. 
This representation is quite sparse since any element not present in the composition is set to zero (and the most number of elements in any given observation is 9).
The values in the vector are normalized by their $L_1$ norm (i.e., the atomic fractions sum to one).
This vector is then used with as an input to a Deep Neural Network (DNN) that consists of multiple fully connected layers, similar to the ElemNet architecture from Ref.~\citenum{Jha2018Elemnet}.
As neurons of fully connected layers connects to every neurons of the previous layer, the order in which the elements are arranged in the input vector has no effect. 
Here we use the alphabetical order to arrange the elements in the representation, though the order is completely arbitrary and simple experiments can be performed to prove that permutations of the element ordering leads to statistically indistinguishable model performance.
The output from the penultimate fully-connected layer will be referred to as the latent code.
The schematic of the 1D unstructured representation and DNN is shown on the left side of Fig.~\ref{fig:representation_schemes}

\subsubsection{1D structured representation}

Like the unstructured representation, the 1D structured representation schemes also uses a fixed vector array of dimension 103 to encode the atomic fractions of elements in a material.
However, as the data will be evaluated with methods that respect spatial structuring, the arrangement of the elements in the representation should follow some principled ordering. 
Here, we have considered several popular one-dimensional ordering schemes:

\begin{itemize}

    \item \textbf{Periodic} - Perhaps the most widely-known chemical order based on the work of Mendeleev in the $19^{\mathrm{th}}$ century, this scheme uses the increasing order of atomic numbers to arrange the elements in the vector. 
    While this results in similar elements being grouped together when using the 2D periodic table, similar elements end up farther away from each other when arranged in a 1D vector.
     
    \item \textbf{Pettifor} - This alternate chemical ordering was proposed by Pettifor\cite{pettifor1984chemical} in 1984 by attempting to place chemically similar elements in neighboring positions in a 1D ordering. 
    This was done by considering the structural stability of 574 binary AB compounds and some additional binary phases.
    He created a structure map by plotting the observed structure for every A and B pair and then ordered the elements to achieve the maximum structural separation in the 2D map and was able to achieve near-perfect separation of the AB binaries. 
    
    \item \textbf{Modified Pettifor} - Inspired by the work of Pettifor, Glawe et al.\cite{Glawe2016modifiedpettifor} performed a statistical study of the likelihood of an element A substituting element B in a crystal structure.
    For this, they used the structural information from the Inorganic Crystal Structure Database (ICSD)  to construct a matrix with each entry (A,B) quantifying how chemically similar two elements are.
    By maximizing the diagonal character of the matrix, they were able to obtain a chemical order similar to the Pettifor scale with some corrections, like the grouping of halogens and magnetic metals.
    By encompassing the information from all available information on crystal structure using a data-driven scheme, this marginally improves the Pettifor order. 
    
    \item \textbf{Arbitrary} - To provide a control for our study and provide a comparison against 1D representations with no meaningful ordering, we also considered the case of arbitrary ordering of elements.
    We use two different arbitrary schemes - an alphabetical ordering (CNN1-alph) and a completely randomized order (CNN1-rand) by shuffling the elements randomly.
    
\end{itemize}

The 1D structured inputs are also normalized to ensure that the atomic fractions sum to 1.
The input vectors are passed through a Convolutional Neural Network (CNN) that consist of multiple 1D convolutional layers and pooling layers, which we refer to as CNN1. 
Unlike fully connected layers, convolutional layers only views small, local regions of the input at a time by using a set of filters with learned weights.
Each of these filters slide across the input along one dimension and learns to extract important spatial correlations in the input.
The pooling layers gradually reduces the spatial size of the convolved features and consequently reduces the number of trainable parameters and complexity of the model.
The output from the last convolutional layer is then flattened to produce the latent code, which represents the features extracted from the 1D structured input.
The latent code is then passed to a fully-connected layer to obtain the output.
The schematic of the 1D structured representation and 1D CNN is shown in the center of Fig.~\ref{fig:representation_schemes}.

\subsubsection{2D structured representation}

Ref.~\citenum{Feng2021} introduced a 2D pseudo-image representation of elemental compositions which they labeled as the Periodic Table Representation (PTR).
In PTR, the alloy compositions are mapped into a fixed 2D vector of dimensions $9 \times 18$ pixels, as can be seen in Fig.~\ref{fig:representation_schemes}.
The yellow cells represent the 108 elements of the periodic table and the atomic fractions of the elements in the HEA compositions are mapped to the corresponding cell in the PTR.
For example, the first cell in the second row in the PTR would be used to store the Li atomic fraction in the HEA composition.
The remaining 54 cells are the unused portion of the periodic table and, along with squares corresponding to the non-occurring elements in the composition, are set to 0.
By comparing against other 2D mappings without periodic table structure like the atom table representation and the randomized periodic table representation, the authors determined that the domain knowledge embedded in PTR provides an advantage over the other schemes for training ML models.

Similar to the 1D structured representations, the 2D PTR input is normalized ($L_1$ norm) and passed through a Convolutional Neural Network (CNN2).
However, the architecture now uses 2D convolutional layers and pooling layers, as can be seen in the right side of Fig.~\ref{fig:representation_schemes}.
The flattened output from the last convolutional layer (the latent code) represents the features extracted from the PTR and is passed to the final fully-connected layer to obtain the output.

\subsection{Prediction of Glass Forming Ability}

From the description of the GFA dataset, it can be understood that the formation of a crystalline or amorphous alloy depends on the alloy composition as well as the processing condition.
However, in the representation schemes and deep learning architectures described above, the elemental compositions are the only input. 
In order to include any additional features required for the deep learning task (like processing parameter for the GFA prediction in this case), they can be appended to the respective latent codes of each model before passing to the final fully-connected layer.
For the GFA predictions, the processing parameters are represented by an additional vector of size 1 with label of 0 for rapid solidification melt-spun or 1 for copper mold casting.
This is a departure from the method of Ref.~\citenum{Feng2021}, wherein processing conditions where included within the pseudo-image itself by mapping them to unused cells in the image, such as the rows above the transition metals.
However, we feel that only composition data should be included in the convolution layers, since these data should not be spatially restricted.
For instance, if the processing conditions are included in the first row and the kernel width is 4, it would require multiple layers to convolve processing conditions with a row 5 element like Zr compared to only one layer for a row 4 element like Ti.

The combination of the different models and input schemes discussed in the previous section result in 7 different models to train. 
All the models have been implemented using \texttt{pytorch} \cite{pytorch2019}. 
The models were trained using batches of size 64 for 2000 epochs using the Adam optimizer and a learning rate of \num{2e-4}.
We also used 10-fold cross validation to determine the average classification performance.
In an attempt to equalize the learning capacity of each model, we kept the number of trainable parameters nearly the same for all models by adjusting hyperparameters like the number of neurons in linear layers and the kernel sizes in the convolutional layers.
The details of the three different kinds of neural networks used in this study are shown in Table.~\ref{tab:model_params}

\begin{table}[ht]

\centering
    \caption{The different models and representation schemes considered in this study along with their respective architectures and number of trainable parameters.}
  \begin{tabular}{|P{1cm}|c|P{5cm}|P{2cm}|}
  
  \hline
     \textbf{Model} & \textbf{Name} & \textbf{Architecture} & \textbf{No. of trainable parameters}  \\
     \hline
     DNN & - &  3 hidden layers with 42 neurons each & 6218\\
     \hline
     \multirow{5}{*}{CNN1} & CNN1-atom & \multirow{4}{5cm}{\centering 3 1D Convolutional layers with [8,16,32] channels and kernel of size 9 + linear layer with 289 neurons} & \multirow{4}{*}{6178}\\
     & CNN1-alph & & \\
     & CNN1-rand & & \\
     & CNN1-pet & & \\
     & CNN-modpet & & \\
     \hline
     \multirow{4}{*}{CNN2} & \multirow{4}{*}{CNN2-PTR} & \multirow{4}{5cm}{\centering 3 2D Convolutional layers with [8,16,32] channels and 3x3 kernel + linear layer with 257 neurons} & \multirow{4}{*}{6146}\\
     & & & \\
     & & & \\
     & & & \\
     \hline
    \end{tabular}
    
  \label{tab:model_params}
\end{table}

\subsection{Transfer learning for prediction of HEA phases, hardness, and yield strength}\label{sec:TL}

As mentioned in the introduction and while describing the HEA phase hardness, and yield strength datasets, the small size of a dataset makes it difficult to train a NN-based model directly.
While the output domains of the GFA and HEA phase/hardness/yield strength prediction tasks are different (binary versus multi-class classification for GFA-HEA phase prediction , classification versus regression for GFA-HEA hardness/yield strength prediction), there is a sufficient overlap between the input domains of these tasks to expect some information from one to transfer to the other.
Additionally, previous ML studies using material attributes as descriptors have shown that there is overlap between the descriptors used in different tasks (like atomic size difference and valence electron concentration in hardness and yield strength predictions) \cite{Bhandari2020,Yang2022}.
Therefore, such situations require alternative strategies like TL to utilize the shared knowledge for improved predictions.

\begin{figure}[h]
    \centering
    \includegraphics[width=\textwidth]{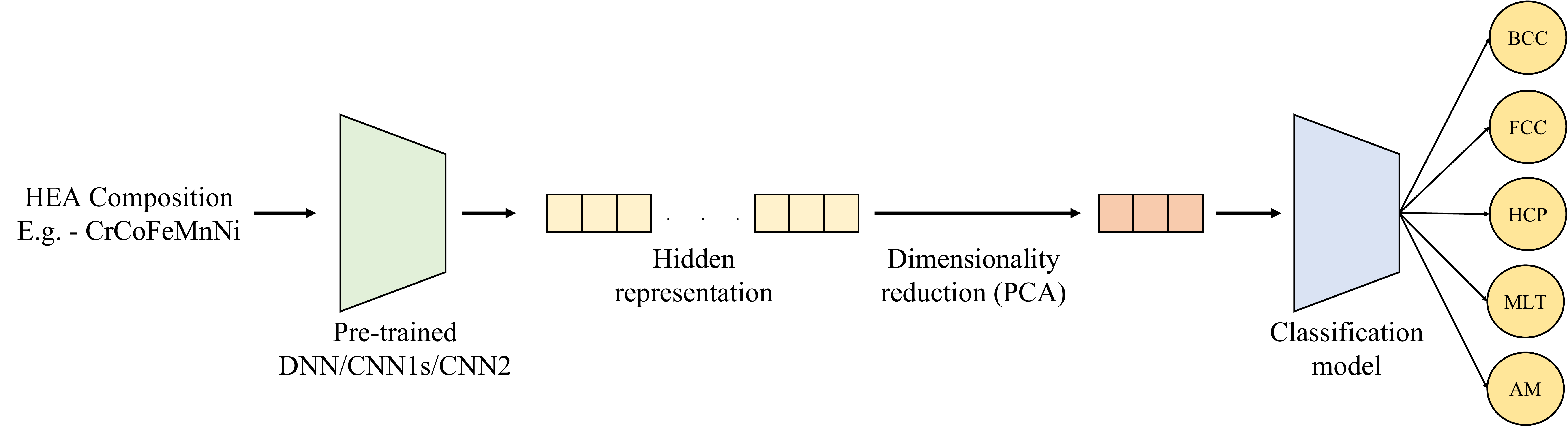}
    \caption{Transfer learning using trained NN models.
    The schematic shows how the TL features can be used for classification of HEA alloys into different phases.
    For regression tasks, the classification model is replaced with a comparable regression model.}
    \label{fig:transfer_learning}
\end{figure}

The TL protocol we used has been adapted from Ref.~ \citenum{Feng2021}, wherein we utilize the different models trained on the GFA prediction task to obtain the latent codes of the HEA compositions. 
As the latent codes are very high-dimensional, dimensionality reduction using Principal Component Analysis (PCA) was performed before applying them to the other tasks.
the PCA embedding of the latent code to obtain the reduced representations (which we will refer to as TL features) were performed  in two systematic ways:
\begin{enumerate}
    \item Using a fixed number of components for all the different latent codes.
    This ensures that the dimensions of the TL features remain small and consistent, even if each type of TL feature contains different amounts of information.
    Here, we use 5, 10 and 20 components in our experiments.
    \item Using the number of components that explain a certain amount of variance in the dataset for each latent code.
    This ensures that each type of TL features has access to the same amount of information.
    Here, we use cutoff values of 90\%, 95\% and 99\% explained variance in our experiments.
\end{enumerate}

The TL features were then as inputs in a ML classfication or regression model, which was chosen to be a \texttt{RandomForestClassifier} (RFC) for the HEA phase prediction task and a \texttt{RandomForestRegressor} (RFR) for the HEA hardness and yield strength prediction task.
The choice was made as Random Forest models were found to outperform several other ML models like Ridge, Lasso, Support Vector Machines, and KNeighbors.  
However, as the yield strength is also a function of the additional features (as-cast/not as-cast, single phase/multi phase, temperature) described previously, they were concatenated with the latent code of the composition for the case of yield strength prediction.
All these models were implemented using the \texttt{sklearn} python package \cite{sklearn}. 
A schematic of the transfer learning protocol is shown in Fig.~\ref{fig:transfer_learning}.

For both the cases, all the model parameters were set to their default values.
While the performance of the models could be improved by performing some hyperparameter tuning, keeping a fixed model allows us to compare the outputs based on the different inputs provided to the model. 
Like before, 10-fold cross validation was performed to determine the average classification and regression performance on the HEA phase and hardness dataset respectively.
We decided to use a NN with the same architecture as described in Table~\ref{tab:model_params} as baseline to compare how NN models perform on such small datasets.
We also used a RFC/RFR model using the atomic fractions of the alloy compositions as input to also evaluate if using TL features instead of these features can provide any advantage in model performance.
  
\subsection{Generalizability test using random subsampling}

Ref~.\citenum{Feng2021} mentioned that the TL features would be beneficial for training ML models on small datasets due to embedding some prior knowledge.
They hypothesized that this will lead to less overfitting, and consequently, better predictions on newer data.
Thus, to investigate the impact of dataset size on the TL model's ability to generalize to new data, we compared the model performance as a function of training data used.
We did this by randomly sampling a subset of the training dataset for training and using the remaining for evaluating the model's prediction. 
We repeated the random subsampling 10 times each for 10 different ratios of splitting the dataset.
Like before, we also include a NN baseline model and a RFC/RFR model using the unstructured representation as input to compare against the RFC/RFR models using the TL features.  

\subsection{Evaluation criteria}
 
For both the classification tasks, the average $F_1$ score from the 10-fold cross validation was chosen as the evaluation metric. 
We have used the implementation of $F_1$ score function included in the \texttt{sklearn} package \cite{sklearn}.
For the regression task, the average root mean squared error (RMSE) from the 10-fold cross validation was used.

We performed statistical significance tests to evaluate if the outputs from the different classifier and regression models result in the same proportions of error or not. 
For both the classification and regression tasks, we used the paired t-test to compare the means of the evaluation metric ($F_1$ score or RMSE) over the 10 folds.
We used the implementation of the paired t-test in the \texttt{scipy} python package \cite{scipy}.
The function returns the $p$ value, which allows us to reject ($p < 0.05$) or support ($p > 0.05$) the null or alternative hypothesis.
For the GFA dataset results, we used the two-tailed t-test with the null hypothesis that two different models have equal proportions of error.
For the transfer learning results, we used the two-tailed t-test with the null hypothesis that the mean performance of the RF models using the latent codes is the same as the mean performance of the baseline NN model trained for only that task.

\section{Results and Discussion}

\subsection{GFA performance}

\begin{figure}[h!]
        \centering
        \includegraphics[width=0.7\textwidth]{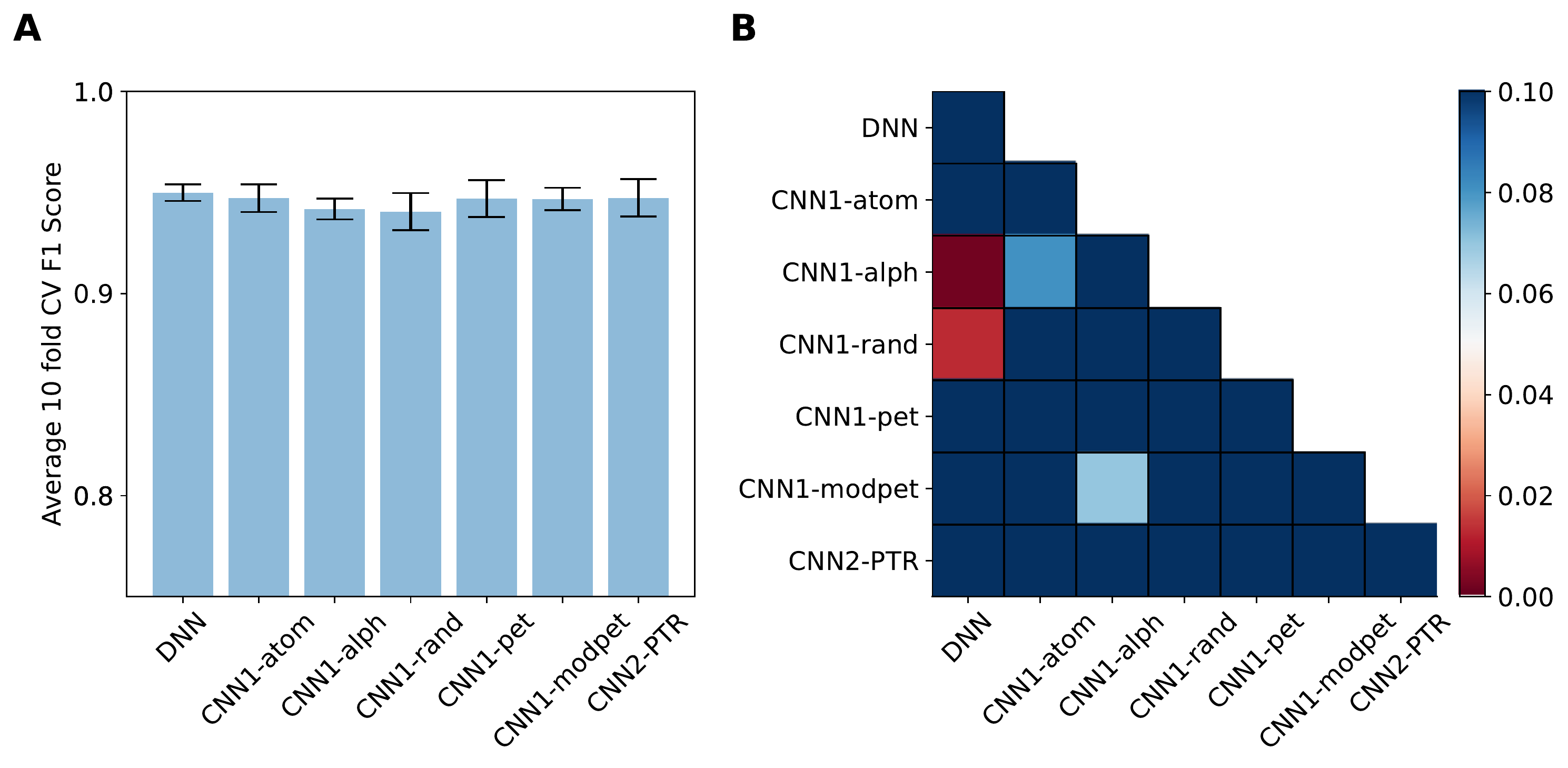}
        \caption{Results from the DNN/CNN training on the GFA dataset.
        A) The average F1 scores from 10 fold CV
        B) $p$-values obtained from two-tailed independent t-test between each combination of the average F1 scores from the DL models.}
        \label{fig:gfa_results}
\end{figure} 

The average $F_1$ scores from the six different deep learning models trained on the GFA dataset listed in Table~\ref{tab:model_params} are shown in Fig.~\ref{fig:gfa_results}A while the $p$-values obtained for each model pair from the two-tailed t-test is shown in Fig.~\ref{fig:gfa_results}B.
The DNN model has the highest average $F_1$ score and very low standard deviation. 
The CNN1-atom, CNN1-pet, CNN1-modpet and CNN2-PTR models have similar $F_1$ scores to the DNN, both the CNN1-rand and CNN1-alph models show the worst performance.
However, the standard deviations for the models using the structured representations is much higher than the DNN using unstructured representation.
This is also apparent from the $p$-values from the two-tailed Student's t-test, which shows that the results from CNN1-alph and CNN1-rand are significantly worse than the results from the DNN, while the null hypothesis cannot be rejected for the other pairs. 
Thus, while we conclude that using an arbitrary spatial ordering leads to worse model performance compared to unstructured representations, we also observe that the structured representation schemes do not provide any significant improvement compared to the DNN model when trained for a specific task.

\subsection{Transfer learning on HEA phases}  

\begin{figure}[h!]
     \centering
     
    \includegraphics[width=0.8\textwidth]{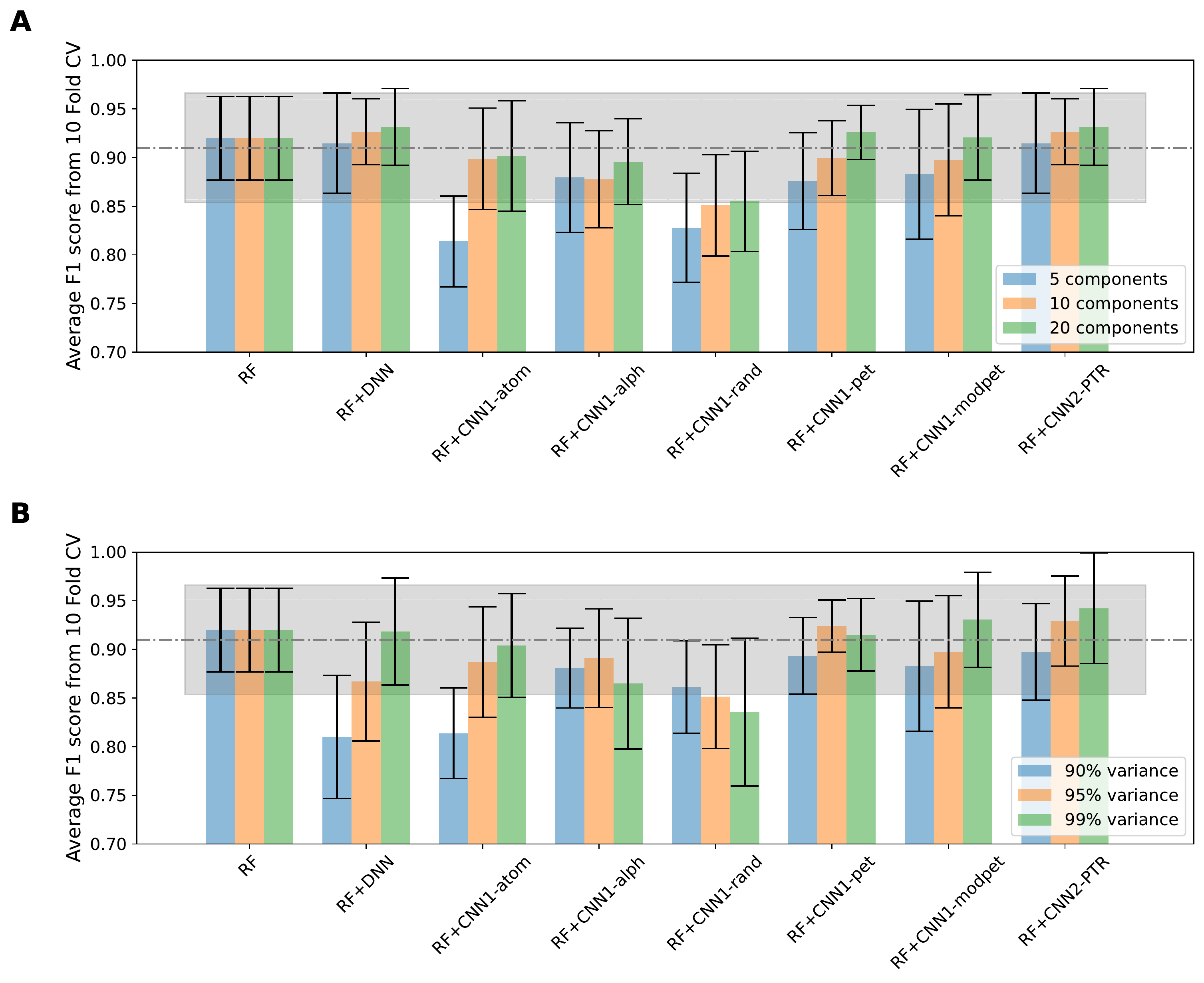}
    \caption{Comparison between the different inputs to the RF models (the alloy composition and TL features from the trained DNN/CNNs) for predicting the HEA phases.
    The grey horizontal line and rectangle represent the mean F1 score and standard deviation in F1 score respectively from a baseline NN model with the same architecture as the DNN used in GFA prediction.
    A) Results for fixed number of components.
    B) Results for fixed explained variance.
    }
        \label{fig:hea_phase_transfer}
\end{figure}

For both the PCA reduction schemes discussed in Section~\ref{sec:TL}, the average $F_1$ scores from the eight RFC models using the different inputs (atomic fraction + the seven different types of TL features) are shown in Fig.~\ref{fig:hea_phase_transfer}.
The RF with atomic fraction as input shows better performance than the NN baseline.
Of the TL features, only the RF+DNN and RF+CNN2-PTR models outperform the baseline for the three different fixed number of components chosen.
The RF+CNN1-pet and RF+CNN1-modpet models outperform the baseline when 20 PCA components are used.
For the fixed variance cases, all the RF using TL features as input are worse than the baseline for the 90\% variance. 
For higher explained variance cutoffs, the RF models using the TL features from DNN, CNN1-pet, CNN1-modpet and CNN2-PTR outperform the baseline.
Interestingly, the average $F_1$ score of the RF+CNN1-alph and RF+CNN1-rand models decrease with higher explained variance, indicating that retaining more information from these arbitrary orderings ends up harming the model's performance.
This suggests that spurious relations retained from the meaningless ordering of the elements confuses the tree model.

\begin{table}[h!]
\caption{The $p$-values from the two-tailed Student's t-test to determine if the RF models are significantly better or worse than the baseline NN model for HEA phase prediction.
The highlighted values indicate the scenarios for which the null hypothesis can be rejected.}
\centering
\begin{tabular}{|l|ccc|ccc|}

\toprule
\multirow{2}{*}{\textbf{Model}} & \multicolumn{3}{c|}{\textbf{Fixed components}} & \multicolumn{3}{c|}{\textbf{Fixed variance}}\\

\cline{2-7}

 &\textbf{5} & \textbf{10} & \textbf{20} & \textbf{90\%} & \textbf{95\%} & \textbf{99\%}  \\
\midrule
RF &  0.672 &  0.672 &  0.672 &  0.672 &  0.672 &  0.672\\
RF+DNN &  0.849 &  0.446 &  0.344 &  \textbf{0.002} &  0.128 &  0.746\\
RF+CNN1-atom  &  \textbf{0.001} &  0.656 &  0.755 & \textbf{0.001} &  0.389 &  0.811\\
RF+CNN1-alph &  0.257 &  0.203 &  0.548 &  0.209 &  0.448 &  0.133\\
RF+CNN1-rand &  \textbf{0.005} &  \textbf{0.029} &  \textbf{0.039} &  0.056 &  \textbf{0.032} &  \textbf{0.028}\\
RF+CNN1-pet & 0.177 &  0.634 &  0.442 &  0.466 &  0.495 &  0.818\\
RF+CNN1-modpet & 0.353 &  0.641 &  0.649 & 0.353 &  0.641 &  0.406\\
RF+CNN2-PTR &  0.849 &  0.446 &  0.344 &  0.610 &  0.427 &  0.232\\
\bottomrule
\end{tabular}
\end{table}

Furthermore, the $p$-values from the two-tailed Student's t-test also suggests that the null hypothesis can be rejected for RF+DNN for 90\% fixed variance, RF+CNN1-atom for 5 fixed components and 90\% fixed variance and RF+CNN1-rand for all cases except the 90\% fixed variance case. 
As these models also show lower average $F_1$ score compared to the baseline NN model, we can conclusively say that these models are significantly worse than the baseline. 

\subsection{Transfer learning on HEA hardness}  

\begin{figure}[h!]
     \centering
         \includegraphics[width=0.8\textwidth]{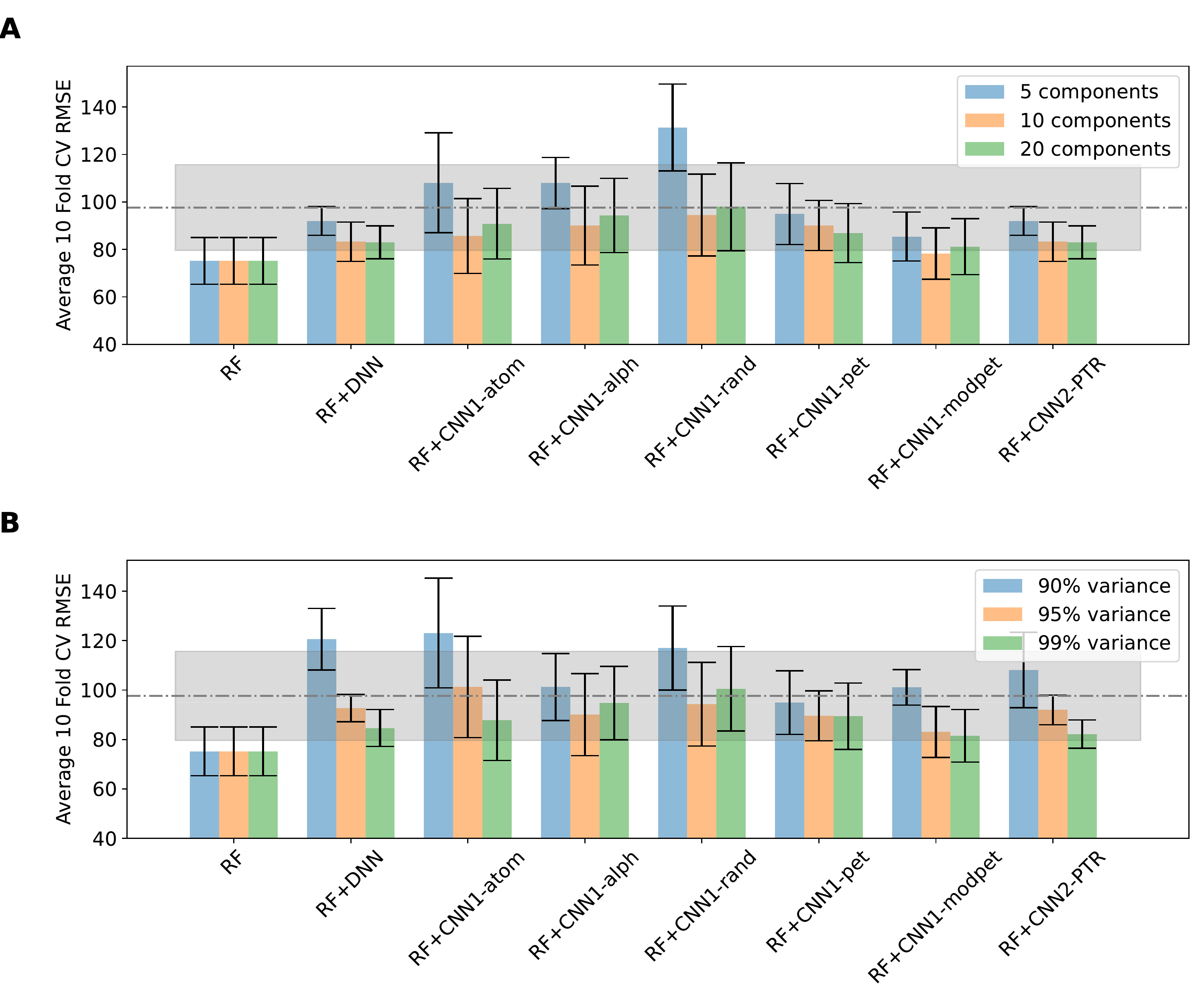}
     
        \caption{Comparison between the different inputs to the RF models (the alloy composition and TL features from the trained DNN/CNNs) for predicting the HEA hardness. The grey horizontal line and rectangle represent the mean RMSE/R and standard deviation in RMSE/R respectively from a baseline NN model with the same architecture as the DNN used in GFA prediction.
        A) Results for fixed number of components.
        B) Results for fixed explained variance.
        }
        \label{fig:hea_hardness_transfer}
\end{figure}

The average RMSE from the eight RFC models using the different inputs (atomic fraction + the seven different types of TL features) for the hardness dataset are shown in Fig.~\ref{fig:hea_hardness_transfer}.
For RMSE, any model with value below the baseline is considered better than the baseline.
Fig.~\ref{fig:hea_hardness_transfer}A shows the results for the keeping the same number of PCA components for all the TL features while Fig.~\ref{fig:hea_hardness_transfer}B shows the results for different number of PCA components for the TL features but cumulatively explaining the same amount of variance in the dataset.

The RF model using atomic fractions as features performs significantly better than the baseline NN and even shows the lowest RMSE.
The RF+DNN, RF+CNN1-pet, RF+CNN1-modpet and RF+CNN2-PTR models show lower RMSE compared to the baseline NN for all the three fixed number of PCA components chosen.
For the fixed variance case, all the RF models using TL features except RF+CNN1-pet are worse than the baseline when number of PCA components are chosen to represent 90\% of the variance in the dataset.
Like before, with increasing the explained variance of the PCA components, the model performance improves.
As was observed for phase prediction, using more components that explain more variance in the dataset end up being detrimental to the model's performance when using the CNN1-rand and CNN1-alph TL features.

\begin{table}[ht]
\centering
\caption{The $p$-values from the two-tailed Student's t-test to determine if the RF models are significantly better or worse than the baseline NN model for HEA hardness prediction.
The highlighted values indicate the scenarios for which the null hypothesis can be rejected.}
\begin{tabular}{|l|ccc|ccc|}

\toprule
\multirow{2}{*}{\textbf{Model}} & \multicolumn{3}{c|}{\textbf{Fixed components}} & \multicolumn{3}{c|}{\textbf{Fixed variance}}\\

\cline{2-7}

 &\textbf{5} & \textbf{10} & \textbf{20} & \textbf{90\%} & \textbf{95\%} & \textbf{99\%}  \\
\midrule
RF &   \textbf{0.005} &  \textbf{0.005} &  \textbf{0.005} & \textbf{0.005} &  \textbf{0.005} &  \textbf{0.005}\\
RF+DNN &   0.389 &  \textbf{0.049} &  \textbf{0.042} &  \textbf{0.006} &  0.446 &  0.069\\
RF+CNN1-atom  &  0.276 &  0.150 &  0.394 &  \textbf{0.016} &  0.699 &  0.238\\
RF+CNN1-alph &  0.162 &  0.364 &  0.677 &  0.642 &  0.364 &  0.712\\
RF+CNN1-rand &  \textbf{0.001} &  0.708 &  0.972 &  \textbf{0.031} &  0.685 &  0.732\\
RF+CNN1-pet &  0.716 &  0.293 &  0.158 &  0.716 &  0.260 &  0.286\\
RF+CNN1-modpet &  0.098 &  \textbf{0.015} &  \textbf{0.036} &  0.607 &  0.053 &  \textbf{0.035}\\
RF+CNN2-PTR & 0.389 &  \textbf{0.049} &  \textbf{0.042} &  0.202 &  0.389 &  \textbf{0.032}\\
\bottomrule
\end{tabular}
\label{tab:p_vals_hardness}
\end{table}

Table.~\ref{tab:p_vals_hardness} shows the $p$-values from the two-tailed Student's t-test on the RMSE score distribution from the eight RF models.
As the null hypothesis can be rejected for the RF model using the atomic fractions as input, and its average RMSE is lower than the baseline, this indicates that this model is significantly better than the baseline.
For the models using the TL features, the RF+DNN and RF+CNN1-atom for 90\% fixed variance, RF+CNN1-rand for 5 fixed components and 90\% fixed variance are significantly worse than the baseline, while RF+DNN for 10 and 20 fixed components, RF+CNN1-modpet and RF+CNN2-PTR for 10 and 20 components and 99\% variance are significantly better.
This again highlights that the CNN1-atom and CNN1-rand TL features are perhaps not as informative as the other structured representation derived TL features.

\subsection{Transfer learning on HEA Yield Strength dataset} 

\begin{figure}
     \centering

         \includegraphics[width=0.8\textwidth]{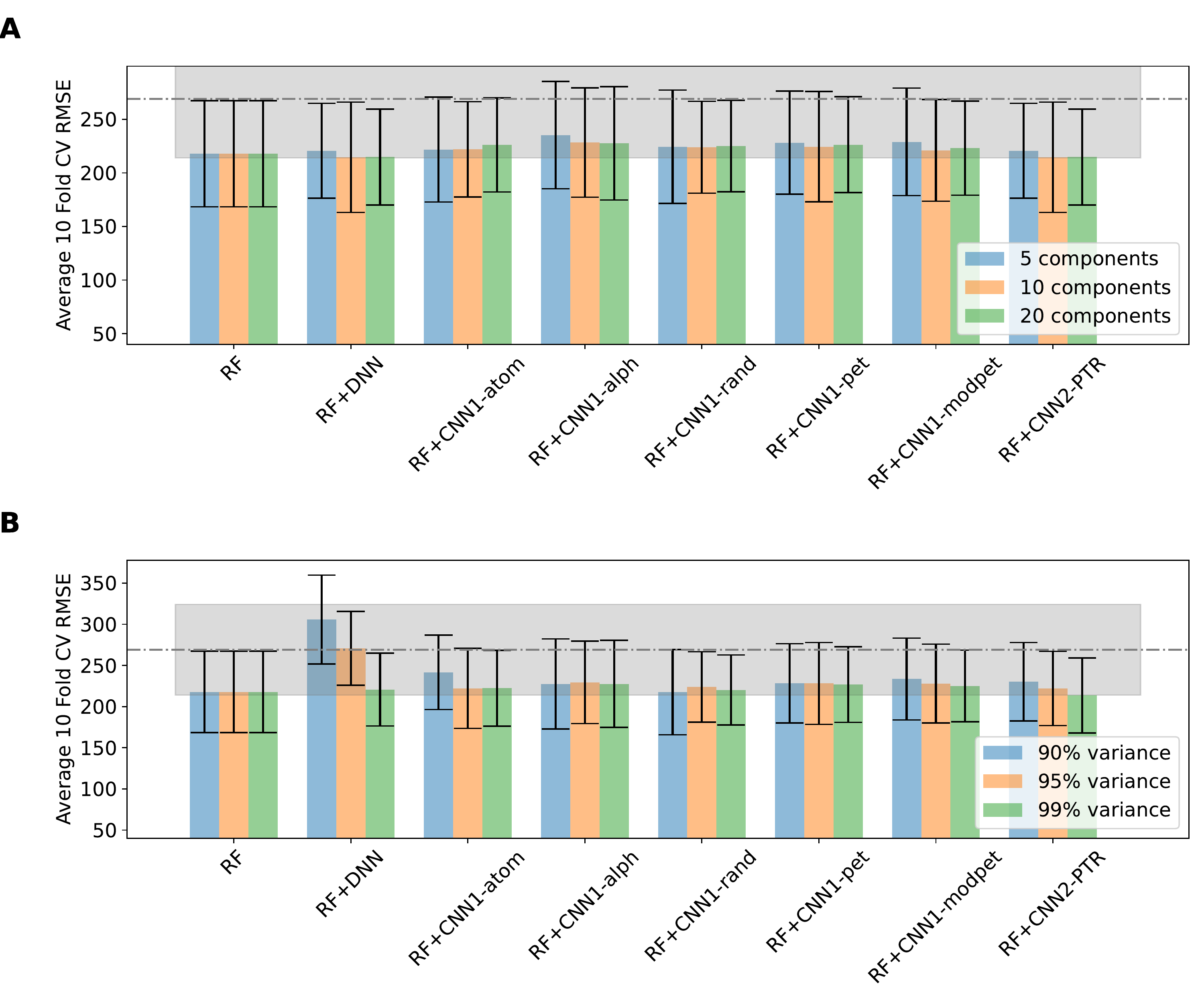}

        \caption{Comparison between the different inputs to the RF models (the alloy composition and TL features from the trained DNN/CNNs) for predicting the HEA yield strength. The grey horizontal line and rectangle represent the mean RMSE/R and standard deviation in RMSE/R respectively from a baseline NN model with the same architecture as the DNN used in GFA prediction.
        A) Results for fixed number of components.
        B) Results for fixed explained variance.
        }
        \label{fig:hea_ys_transfer}
\end{figure}

The average RMSE from the eight RFC models using the different inputs (atomic fraction + the seven different types of TL features) for the yield strength dataset are shown in Fig.~\ref{fig:hea_ys_transfer}.
Fig.~\ref{fig:hea_ys_transfer}A and Fig.~\ref{fig:hea_ys_transfer}B respectively show the results for the keeping the same number of PCA components and different number of PCA components but explaining the same amount of variance in dataset.
Rather surprisingly, we can observe that all the RFR models perform better than the baseline NN model for the different instances of fixed number of PCA components. 
For the case of different variances, only the RF+DNN model shows inferior performance compared to the baseline NN model, that too only for 90\% variance.
Surprisingly, the two random order utilizing models (RF+CNN1-alph and RF+CNN1-rand) both show better performance than the baseline, and even shows comparable performance to the RFR models using the structured representation derived TL features.
This is also supported by the $p$ values obtained from the Student's t-test test shown in Table.~\ref{tab:p_vals_ys}.
Of all the models, only the RF+DNN for 10 and 20 components and RF+CNN2-PTR for 10 and 20 components and 99\% variance are significantly better than the baseline NN model.\\

\begin{table}[ht]
\caption{The $p$-values from the two-tailed Student's t-test to determine if the RF models are better than the baseline model for HEA Yield Strength prediction.
The highlighted values indicate the scenarios for which the null hypothesis can be rejected.}
\centering
\begin{tabular}{|l|ccc|ccc|}

\toprule
\multirow{2}{*}{\textbf{Model}} & \multicolumn{3}{c|}{\textbf{Fixed components}} & \multicolumn{3}{c|}{\textbf{Fixed variance}}\\

\cline{2-7}

 &\textbf{5} & \textbf{10} & \textbf{20} & \textbf{90\%} & \textbf{95\%} & \textbf{99\%}  \\
\midrule
RF & 0.052 &  0.052 &  0.052 &  0.052 &  0.052 &  0.052 \\
RF+DNN & 0.055 &  \textbf{0.044} &  \textbf{0.034} & 0.170 &  0.943 &  0.055 \\
RF+CNN1-atom &  0.070 &  0.062 &  0.084 &  0.265 &  0.072 &  0.067 \\
RF+CNN1-alph &  0.188 &  0.121 &  0.120 &  0.126 &  0.128 &  0.120\\
RF+CNN1-rand &  0.096 &  0.069 &  0.075 &  0.056 &  0.069 &  0.051\\
RF+CNN1-pet &  0.111 &  0.093 &  0.088 &  0.111 &  0.115 &  0.095\\
RF+CNN1-modpet & 0.123 &  0.063 &  0.066 & 0.168 &  0.109 &  0.077\\
RF+CNN2-PTR &  0.055 &  \textbf{0.044} &  \textbf{0.034} & 0.127 &  0.063 &  \textbf{0.032} \\
\bottomrule
\end{tabular}
\label{tab:p_vals_ys}
\end{table}

\subsection{Random subsampling}  

\begin{figure}
     \centering
         \includegraphics[width=0.9\textwidth]{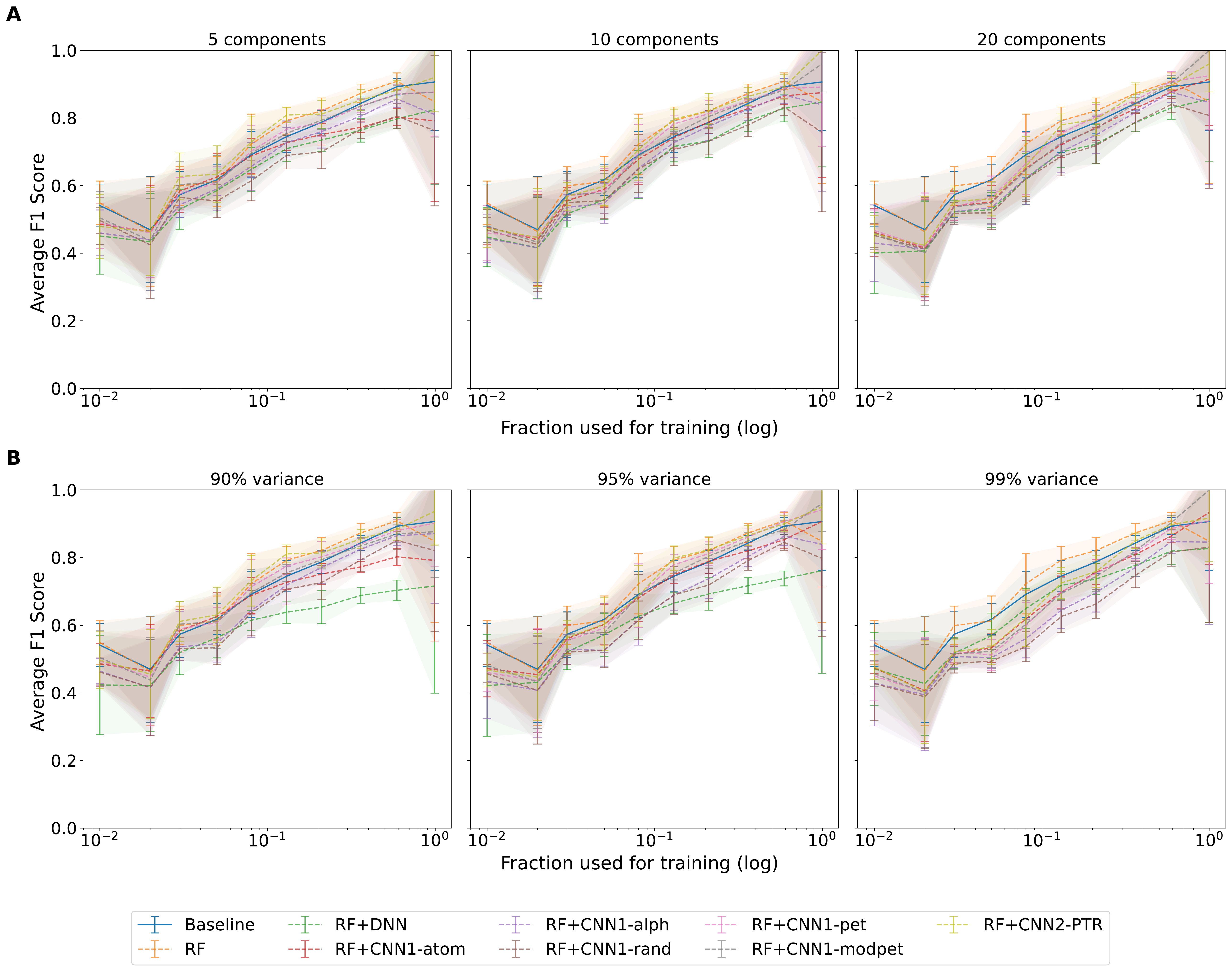}
     
        \caption{The $F_1$ score of the different models as a function of the amount of training data used for the HEA phase prediction task.
        A) Results for fixed number of components.
        B) Results for fixed explained variance.}
        \label{fig:hea_phase_general}
\end{figure}

\begin{figure}
     \centering
         \includegraphics[width=0.9\textwidth]{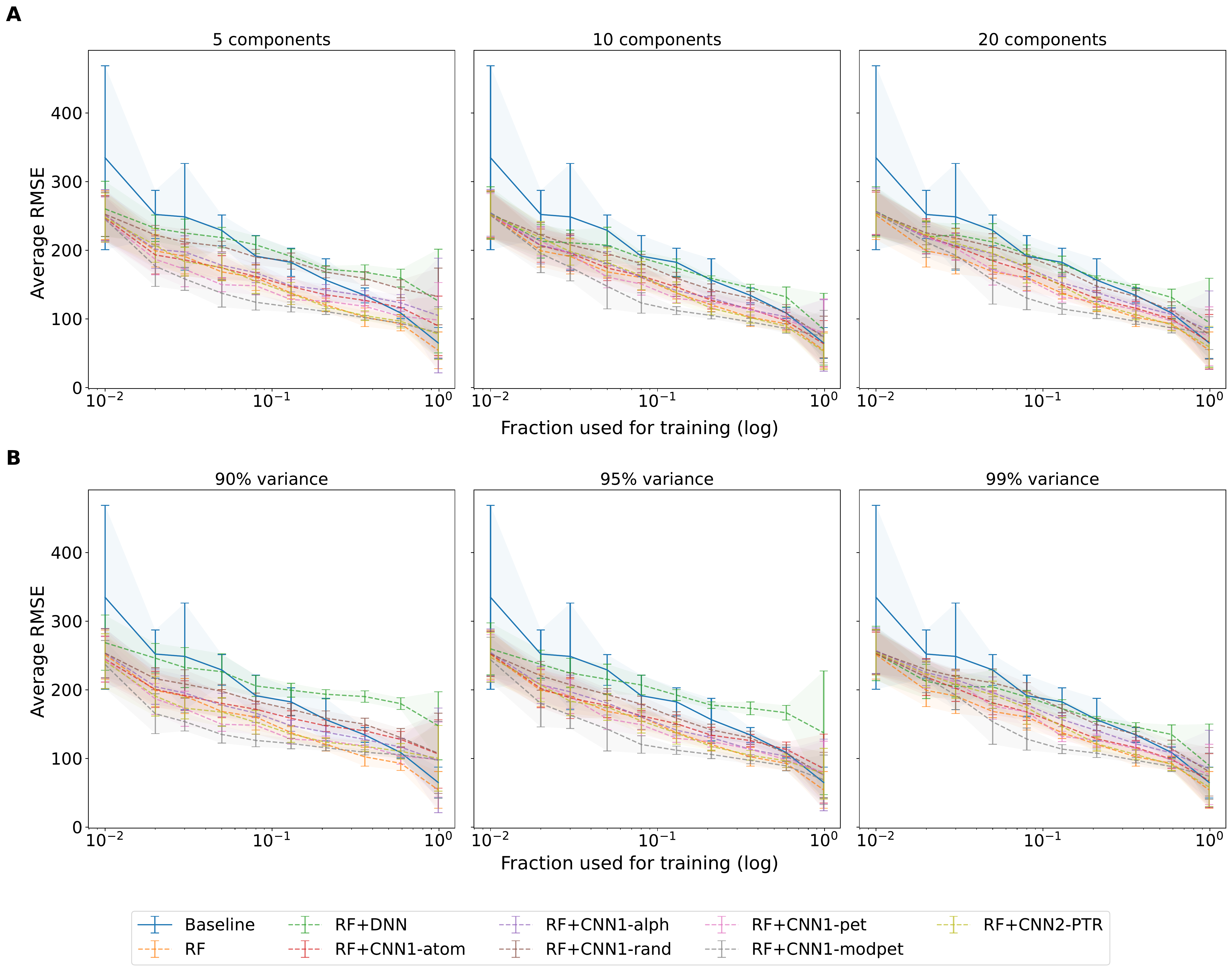}
        \caption{The RMSE of the different models as a function of the amount of training data used for the HEA hardness prediction task.
        A) Results for fixed number of components.
        B) Results for fixed explained variance.}
        \label{fig:hea_hardness_general}
\end{figure}

\begin{figure}[h!]
     \centering
         \includegraphics[width=0.9\textwidth]{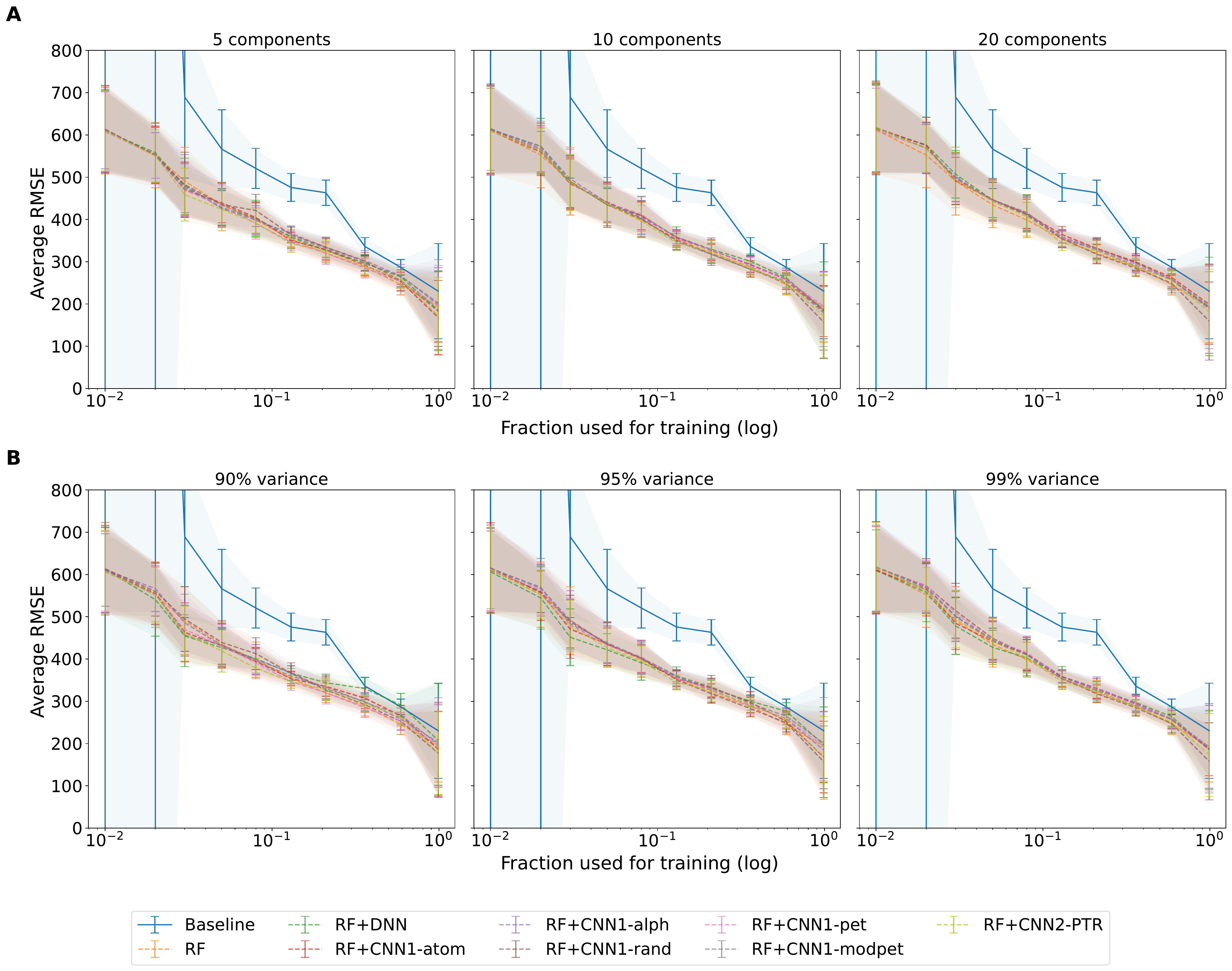}
     
        \caption{The RMSE of the different models as a function of the amount of training data used for the HEA yield strength prediction task.
        A) Results for fixed number of components.
        B) Results for fixed explained variance.
        For the first two observations, the RMSE for the baseline NN model is greater than 1000.}
        \label{fig:hea_ys_general}
\end{figure}

We selected 10 different ratios of splitting the entire dataset using a log scale.
This was done to evaluate the models more in the scenarios corresponding with lower volume of training dataset. 
The results from the 10 times random subsampling for the HEA phase, hardness and yield strength datasets for both fixed components and fixed variances are shown in Fig.~\ref{fig:hea_phase_general}, Fig.~\ref{fig:hea_hardness_general}, and Fig.~\ref{fig:hea_ys_general} respectively.
Once again, we used the NN model with the same architecture as the DNN in Table~\ref{tab:model_params} as the baseline.
In the case of HEA phase prediction, we observed that most of the RF models actually perform worse than the baseline, with the worst offenders being the models using the DNN and CNN1-rand TL features.
For the case of HEA hardness prediction, we observe that for smaller fractions of the training data used, the RF models using the TL features perform better than the baseline model.
Once again, the RF+DNN and RF+CNN1-rand models show the worst performance out of all the considered RF models. 
For the HEA yield strength dataset, the baseline model is shockingly bad, and consequently all the RF models are much better than the baseline.
However, it should also be noted that the RF using the atomic fractions as input consistently does the best out of all the RF models

From all the tests performed, we observed that the randomly ordered structured representations and their derived TL features are generally worse than the chemically intuitive structured representations and their counterparts, both in terms of TL and generalizability towards new data, while the unstructured representations and their derived TL features are worse when it comes to generalizing to new data.
This is hypothesized to be due to inclusion of spurious information in the arbitrarily ordered structured representations and the lack of any embedded chemical knowledge in the unstructured representation.
Comparatively, the TL features from CNN1-pet, CNN1-modpet and CNN2-PTR generally show the best results compared to the models using the remaining TL features.
However, we also observed that a RF model using atomic fractions of the compositions as input is able to perform comparatively or better than the best RF model using the TL features.

\begin{figure}[h!]
        \centering
        \includegraphics[width=0.8\textwidth]{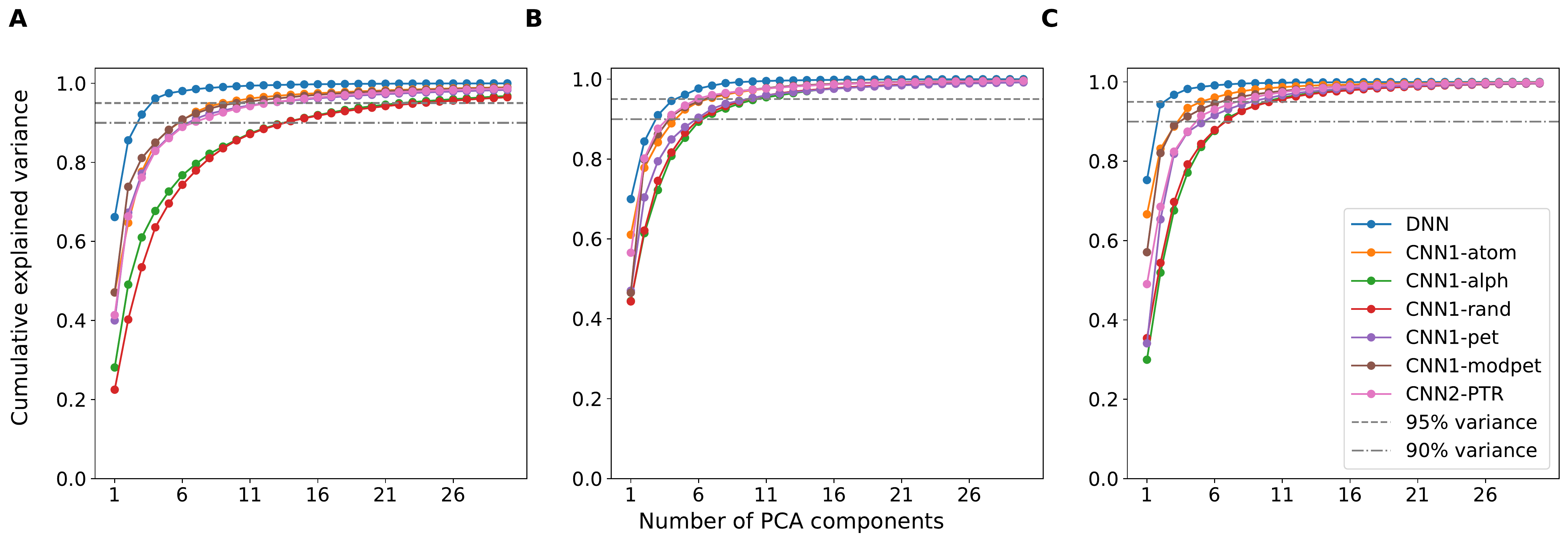}
        \caption{The first 30 PCA components for the TL features along with the cumulative explained variance for a) HEA phase, b) HEA hardness, and c) HEA yield strength datasets.}
        \label{fig:pca_components}
\end{figure}

To further understand to why TL features from DNN, CNN1-rand, and CNN1-alph show the generally worse performance for HEA property prediction, we show the first 30 PCA components for TL features for the three datasets in Fig.~\ref{fig:pca_components}.
In all three cases, the latent codes from the DNN require the least number of components, while the latent codes from CNN1-alph and CNN1-rand require the most number of components. 
The arbitrary 1D ordering means the 1D CNN has a harder time finding useful patterns, while the fully connected layers in the DNN is able to learn weights easily for the neurons that govern each individual element of the unstructured representation.
However, in doing so, the DNN is effectively overfitting to the specific task at hand while losing chemical information that may benefit the transfer learning task.
Within the 1D structured representations, CNN1-pet and CNN1-modpet latent codes need lower number of PCA components than the CNN1-atom.
This indicates that the chemical similarity based grouping of elements in the Pettifor and modified Pettifor orders allows the 1D CNN to extract the relevant information compared to the periodic order or the random order.

Another interesting observation is that the while performance of RF models using CNN1-alph and CNN1-rand TL features on the HEA phase dataset was among the worst compared to other, their performance is comparatively better on the other two predictive tasks. 
We believe that that might be due to lower variability in the HEA hardness and yield strength datasets as compared to the HEA phase dataset.
To verify this, we calculated the variance of the compositions in these datasets from the `mean' composition of that dataset.
The mean composition here refers to the equiatomic composition formed by using all the elements of the specific dataset. 
The variance is then calculated using the atomic fractions of the mean composition and the dataset compositions.
Compared to the variance of 0.224 for the HEA phase dataset, the variance in the HEA hardness and yield strength datasets are 0.119 and 0.116 respectively.
Thus, for datasets with higher variability, the shortcoming of the random ordering -- and consequently the derived TL features -- become more apparent.

\section{Conclusion}

In this paper, we systematically compared some different unstructured and structured representation schemes for HEA chemical compositions.
We began by training deep learning models for each representation type on a classification task using a large dataset of metallic glasses.
The learned representations from these NNs were then reduced to a lower dimension using PCA and used as input features for RF models for predictive tasks on smaller datasets, following the example of Ref.~\citenum{Feng2021}.
We also evaluated if these TL features are more resistant to overfitting and result in better performance by performing a random subsampling of the dataset to make training datasets of different sizes.

The results from these studies indicate that while using arbitrary ordering is detrimental to DNN model performance, there is no clear benefit of using structured representations over a simple unstructured representation when training for a single task as these models shows very similar performance.
For the transfer learning tasks, the TL features from all the chemically meaningful structured representations generally outperform the TL from the unstructured or arbitrary representations, indicating that the embedded structural information preserves relevant domain knowledge.
This is also observed in the generalizability tests, wherein the TL features derived from unstructured and arbitrary arrangements usually perform worse than the baseline (single-task) NN model. 

However, we note that a single-task tree model (Random Forest) using atomic fractions as input was able to show comparable or superior performance in every test, indicating that using such sophisticated transfer learning protocol may not have any benefit.
Ultimately, this may support the widespread notion that tree based models are better suited for tabular data than NNs \cite{grinsztajn2022tree}.
One caveat to this is that tree-based architectures are generally used for discriminative tasks rather than generative.
However, a recent paper\cite{watson2022smooth} discusses a novel method of synthetic tabular data generation using a recursive, adversarial variant of unsupervised random forests.
Therefore, a possible future research direction could be to compare these representation schemes for data generation.
Another interesting direction could also be to evaluate if the recursive RF based generative models outperform traditional NN-based architectures like Generative Adversarial Networks and Variational Autoencoders when used for HEA generation.
Such studies could provide further guidance in selecting the best representations and model architecture for HEA design.

\section{CRediT authorship contribution statement}
AD: Conceptualization, Methodology, Software, Data analysis, Data Visualization, Data Interpretation, Writing – original draft.
WFR: Conceptualization, Supervision, Funding acquisition, Project administration, Writing – review \& editing.

\section{Declaration of Competing Interest}
The authors declare that they have no known competing financial interests or personal relationships that could have appeared to influence the work reported in this paper.

\section{Acknowledgments}
The present work is based upon work supported by the Department of Energy / Advanced Research Projects Agency – Energy (ARPA-E) under award No DE-AR0001435.
The authors thank Adam Krajewski, Shuang Lin, Marcia Ahn, John Shimanek, Lavanya Raman, Wenjie Li, Shunli Shang, Douglas Wolfe, Allison Beese, and Zi-Kui Liu for helpful discussions related to HEA modeling and design.

\section{Software and data availability}
The datasets and code used to generate the results in this work are available at \href{https://github.com/dovahkiin0022/representations}{github.com/dovahkiin0022/representations}


\bibliography{references}  






\end{document}